\documentclass[12pt]{article}\usepackage[hyperfootnotes=false]{hyperref}
\usepackage{epsfig}
\usepackage{float}
\usepackage{empheq}

\usepackage{caption}

\usepackage{amsmath}
\usepackage{amssymb}
\usepackage{graphicx}
\newlength{\abstractwidth}
\renewcommand{\thefootnote}{\fnsymbol{footnote}}
\renewcommand{\thanks}[1]{\footnote{#1}} 
\newcommand{\starttext}{
\setcounter{footnote}{0}
\renewcommand{\thefootnote}{\arabic{footnote}}}

\newcommand{\be}{\begin{equation}}
\newcommand{\bea}{\begin{eqnarray}}
\newcommand{\eea}{\end{eqnarray}}
\newcommand{\beq}{\begin{equation}}
\newcommand{\ee}{\end{equation}}

\newcommand*\widefbox[1]{\fbox{\hspace{2em}#1\hspace{2em}}}

\def\simleq{\; \raise0.3ex\hbox{$<$\kern-0.75em
\raise-1.1ex\hbox{$\sim$}}\; }
\def\simgeq{\; \raise0.3ex\hbox{$>$\kern-0.75em
\raise-1.1ex\hbox{$\sim$}}\; }

\def\bi{\begin{itemize}}
\def\ei{\end{itemize}}

\def\sc{\setcounter{equation}{0}}

\def\bsub{ \begin{subequations}
\begin{empheq}[box=\widefbox]{align}  }
\def\esub{ \end{empheq}
\end{subequations}}

\def\bn{\bigskip \noindent}

\makeatletter
\g@addto@macro\normalsize{%
  \setlength\abovedisplayskip{10pt}
  \setlength\belowdisplayskip{20pt}
  \setlength\abovedisplayshortskip{10pt}
  \setlength\belowdisplayshortskip{20pt}
}
\makeatother

\usepackage{color}

\begin{document}

  
\begin{titlepage}

\rightline{}
\bigskip
\bigskip\bigskip\bigskip\bigskip
\bigskip

\centerline{\Large \bf { Electromagnetic Memory }}
\bn

\bigskip
\begin{center}
\bf      Leonard Susskind  \rm

\bigskip
 Stanford Institute for Theoretical Physics and Department of Physics, \\
Stanford University,
Stanford, CA 94305-4060, USA \\

\end{center}

\bn

\begin{abstract}

An elementary derivation of the electromagnetic memory effect is given. An
experimental setup to detect it is suggested.

\end{abstract}

\end{titlepage}

\starttext \baselineskip=17.63pt \setcounter{footnote}{0}

\Large
Classical memory effects and their relation to BMS conservation laws and soft emission
theorems have been the subject of recent interest by Strominger and collaborators (see   \cite{Kapec:2015ena}
and references contained therein.) In this note I will give an elementary derivation of the
electromagnetic memory effect and suggest a way of detecting it.

\sc
\section{Memory Effect}

Consider a large sphere $\Omega$ surrounding an explosion which ejects charged particles, which
later pass through the sphere. 
We assume the explosion is near the center of the sphere
so that the particles’ velocities are radial when they pass through the sphere. We also
assume that they move with velocity close to or at the speed of light. Before the explosion
the charge density, current density, and electromagnetic fields were zero. We work in the
temporal gauge and take the initial value of the vector potential to be 
$A_{in}=0.$

The Gauss equation
\be
\nabla\cdot E = \rho
\ee
is true everywhere at all time. We will consider it on the sphere.
Since the charges are moving in a lightlike radial trajectory when they pass the sphere
we can assume that the charge density is equal to the radial component of the current $j_r$.

Thus, on the sphere we may write,
\be 
\nabla \cdot \dot{A} = -j_r
\label{1.2}
\ee

where we have used 
\be 
E=-\dot{A}
\label{1.3}
\ee

Now integrate over time and we find (on the sphere) that after all charges have passed
through the sphere,
\be 
\nabla\cdot A = -Q(\Omega, \infty)
\label{1.4}
\ee
where $Q(\Omega, \infty)$ is the total charge that has passed through the point $\Omega$ after all particles
have left.

It is easy to show that the contribution of the normal components of $E$ average to zero
as the charges pass through the surface of the sphere. As the charge recedes from the
surface the normal component of E is opposite to the value it had while the charge was
approaching the surface.
Therefore we may restrict the divergence of $\dot{A}$ 
to the components along the sphere.
The subscript $\Omega$ indicates the restriction to the sphere. Thus at the end of the process we
find that at every point on the sphere:
\be 
\nabla_{\Omega}\cdot A_{\Omega} = - Q(\Omega, \infty)
\ee

Now let us imagine that the sphere is covered with a collection of superconducting
nodes. Initially before the explosion the superconducting nodes are connected by superconducting wires so that the 
relative phases of the superconducting condensates at the nodes are all zero. Then we disconnect the
wires.
At the end of the experiment there is a gauge field $A$ present on the sphere but no
elecrtic or magnetic field. Therefore the gauge field has the form,
\be 
A=\nabla \lambda
\ee

We can eliminate the gauge field by a gauge transformation on the sphere at the cost
of creating a relative phase between the superconducting nodes. The position dependent
phase is just $\lambda$ This frozen-in phase is the electromagnetic memory effect. 

The relative phases of the nodes can be detected by connecting pairs of nodes with Josephson junctions. Josephson currents  
 will flow proportional to the
phase differences.

\sc
\section{Local Conservation Law}
We can express the memory effect as an instantaneous conservation law. Define 
$Q(\Omega, t)$
to be the total charge that has passed through the point $\Omega$ up to time $t.$ Obviously
\be 
\dot{Q}(\Omega, t) = j_r(\Omega, t)
\ee
The Gauss condition becomes,
\be 
\frac{d}{dt} \{ \nabla_{\Omega}\cdot A_{\Omega} + Q(\Omega, t) \} = 0
\label{2.2}
\ee
Equation \ref{2.2}  is a conservation law that is true at every point on $Omega.$ The reason that
it is not trivial is that when integrated over time the change in A is not zero since it
must satisfy \ref{1.4}. As we have seen this leads to an observable flow of charge between
superconductors. The flow will occur when we reconnect the nodes no matter how long
we wait. The memory of the explosion is frozen into the relative phases.
A last point is that the integrated conservation law may also be understood as the
usual soft photon emission theorem. Thus we see the triangle of ideas: Local conservation
law, soft theorem, memory effect.

\sc
\section{Generalization}

The motion of charges does not have to be light-like to have a memory effect although
the analysis is not as elegant. For simplicity assume the charges move radially but this is
not essential. Also assume that the charge density and electromagnetic fields are all zero
inside the sphere at the beginning and end of the process.
The simplest case is when the charges move with fixed velocity $v<1.$ In that case we
can write,
\be 
j_r = v\rho
\ee
and replace \ref{1.2} by
\be 
\nabla\cdot \dot{A} = - \frac{1}{v} j_r
\ee
and \ref{1.3} by,
\be 
\nabla \cdot A =-\frac{1}{v} Q(\Omega, \infty)
\label{3.3}
\ee

If the velocity at the sphere is time dependent then \ref{3.3} becomes more complicated with
the right side being an integral.

  \section*{Acknowledgements}
  
This note was originally a private response to a talk at Stanford by Andy Strominger. I
thank Andy for calling my attention to the memory effect and especially for encouraging
me to publish the note.

This work was supported in part by National Science Foundation grant 0756174 and by
a grant from the John Templeton Foundation. The opinions expressed in this publication
are those of the author and do not necessarily reflect the views of the John Templeton
Foundation.

\end{document}